\documentstyle[preprint,revtex,eqsecnum]{aps}

\draft
\tightenlines
\begin{document}

\preprint{HD--THEP--92--35/R}
\begin{title}
\bf SEMILEPTONIC INCLUSIVE B--DECAYS AND THE\\
DETERMINATION OF $|V_{ub}|$ FROM QCD SUM RULES
\end{title}
\author{Patricia Ball}
\begin{instit}
Physik-Department, TU M\"unchen, D-8046 Garching, FRG
\end{instit}
\author{V.M.~Braun\cite{origin}}
\begin{instit}
Max-Planck-Institut f\"ur Physik, P.O.\ Box 40 12 12, D-8000
M\"unchen 40, FRG
\end{instit}
\author{H.G.~Dosch}
\begin{instit}
Institut f\"ur Theoretische Physik, Universit\"at Heidelberg,
D-6900 Heidelberg, FRG
\end{instit}
\receipt{}
\begin{abstract}
We calculate the electron spectrum of semileptonic decays of
B-mesons
into non-charmed hadrons. The shape of the spectrum obtained from
QCD
sum rules is in general agreement with quark model calculations. At
high electron energies the decrease of the spectrum is less steep
than
predicted by Altarelli et al. Our analysis yields a total decay rate
$\Gamma(B\to X_u e\bar\nu) =  (6.8\pm 2.0)\, 10^{13}\,|V_{ub}|^2\,
\mbox{s}^{-1}$. For the spectrum integrated over electron energies
in the
interval $2.4\mbox{\ GeV}\leq E_e \leq 2.6\mbox{\ GeV}$ in the
laboratory
frame we obtain
$\Gamma(B\to X_u e \bar\nu)= (5.0\pm 1.6)\,10^{12}\, |V_{ub}|^2\,
\mbox{s}^{-1}$. Estimating all sorts of theoretical uncertainties,
we
obtain $|V_{ub}| = 0.003\pm 0.001$ for the $b \to u$ weak
transition
matrix element from the new CLEO-data.
\end{abstract}

\narrowtext

\section{Introduction}

Inclusive semileptonic decays of B-mesons have been studied
\cite{stone,cleo} for already more than a decade. Lacking a reliable
identification of the hadrons in the final state, the data on
semileptonic
decays provide at present the most accurate
information on the determination of the CKM-matrix-element
$|V_{ub}|$. The accuracy of the determination of $|V_{ub}|$ is,
however, limited to a great extent by the fact that the
$b\rightarrow u$ transitions can experimentally be separated from
the
dominating $b\rightarrow c$ decays only
in a rather narrow kinematical region of high electron energies,
close to
the kinematical threshold.

Most theoretical predictions for the spectrum are based on a
spectator model \cite{accm}. This approach makes use of the fact
that
in the limit of infinite b-quark mass $m_b$ the decay of the meson
is
reduced to the $\beta$-decay of a free
b-quark, which at large electron energies is modified by
Sudakov-type radiative corrections. In practice, however, the
b-quark
is not heavy enough, and both the interaction with the
spectator quark and hadronization effects cannot be neglected
even for the total semileptonic decay rate. The numerical importance
of $1/m_b$ corrections becomes apparent already from simple phase
space considerations: since the total decay rate is proportional to
$m_B^5$, the replacement of the hadron by the quark mass
reduces the decay rate by nearly a factor two. Even for the
ratio of semileptonic to hadronic width, where the dimensionful
factor
$m_B^5$ cancels, a calculation of the $b\to c$ transition using
free quarks yields a value which is (10-15)\% below the data
\cite{bjorken,totalrate}. The situation gets considerably worse for
 the
electron spectrum near the charm threshold, where the heavy quark
expansion breaks down and there is no reason to believe that
Sudakov-type effects provide the most important correction. In
existing calculations of the spectrum in this region the deviations
{}from a simple free quark decay picture are taken into account in a
model-dependent way, introducing constituent quark masses and
Fermi-motion of the quarks inside the hadrons by a certain smearing
 of
the momentum \cite{accm} or by using the Peterson fragmentation
function \cite{paschos}. The results show that the influence of the
spectator quark and
hadronization effects become indeed important at large electron
energies, and quantitative predictions depend strongly on the model
assumptions and on the parameter values. Especially unpleasant is
that the kinematical endpoint is not well defined, since it depends
 on
the constituent quark masses used in the model and not on the mass
of
the hadrons.

In an alternative approach advocated in Refs.~\cite{BWS,KS,GISW}
the
electron spectrum of the inclusive decay is obtained by an explicit
summation of contributions of various exclusive channels, e.g.\
$B\to
\pi e \nu$, $B\to \rho e \nu$, which are calculated in the framework
of a  quark model. In this method, the results rely heavily on the
assumed shape of the exclusive form factors. In addition, it is not
possible to take into account exclusive decays with two or more
nonresonant mesons (e.g.\ two pions in an S-state), which, in
contrast
to the decays to charmed final states, are expected to play a
nonnegligible role \cite{RDB}. Recently, a two-component model
has been suggested in Ref.~\cite{RDB}, which aims to combine the
positive and avoid the negative features of both approaches.
However,
 the justification of this two-component model has been questioned
in
Ref.~\cite{isgur}.

In view of conflicting results in the current literature, we find
it
important to address the problem from a completely different
point of view. In this paper we calculate the electron spectrum of
the
inclusive semileptonic B-decay within the framework of QCD sum
rules.
This method is due to Shifman, Vainshtein and Zakharov \cite{SVZ}.
It
is essentially a matching procedure between the operator product
expansion for a suitable correlation function in the not so deep
Euclidean region and the representation of the same correlation
function as dispersion integral in terms of contributing hadronic
states.
Nonperturbative effects are taken into account by nonvanishing
vacuum
expectation values of gauge invariant operators, the so-called
condensates. QCD sum rules have proved to be a very successful tool
 for
calculations of both static and decay properties of light and heavy
hadrons. Although they often provide less detailed results
than quark model calculations, they have the advantage of being
 closer
to the first principles of QCD. In recent papers we have applied
 this
method to the study of exclusive semileptonic decays
\cite{BBD,Btopion,ball}. The application to inclusive decays poses
some new problems, which are discussed below. We show that it is
possible to use the sum rule technique in order to calculate the
electron spectrum up to an electron energy $E_e \simeq
2.1\,\mbox{GeV}$ in the rest frame of the B-meson, which is slightly
 below the charm threshold. The
shape of the spectrum and a constraint at the endpoint, however,
 allow
an extrapolation to the experimentally accessible region
$E_e \agt 2.3\,\mbox{GeV}$ without introducing a large
numerical uncertainty. Predictions for the shape of the spectrum
and
integrated rates are given and compared to other model calculations.

Our paper is organized as follows. In Sec.~II we recollect necessary
kinematical formulae.  In Sec.~III we derive sum rules for
the electron spectrum and discuss their region of applicability,
 which
is constrained by the requirements of a  sufficiently large interval
of duality and the existence of the short-distance expansion. In
Sec.~IV we discuss the heavy quark limit of our sum rules, Sec.~V
contains numerical results and conclusions. Technical details of
the
calculation are given in the appendix.

\section{Kinematics}

The total width $\Gamma$ of the inclusive decay $B\to X_u e^-
\bar\nu$, where $X_u$ denotes a charmless hadronic state, can
be written in terms of a leptonic tensor $L_{\mu\nu}$ and a
hadronic
tensor $W_{\mu\nu}$ as
\begin{equation}\label{eq:gamma}
\Gamma = \frac{G_F^2 |V_{ub}|^2}{32\pi^5 m_B}\int\!\!
\frac{d^3p_e}{2E_e}\, \frac{d^3p_\nu}{2E_\nu}\, W_{\mu\nu}\,
L^{\mu\nu}
\end{equation}
with
\begin{equation}
L^{\mu\nu} = 2 \left( p_e^\mu p_\nu^\nu + p_\nu^\mu p_e^\nu -
g^{\mu\nu}
p_e p_\nu + i \epsilon^{\mu\nu}_{\phantom{\mu\nu}\rho\sigma}p_e^\rho
p_\nu^\sigma \right)
\end{equation}
and
\begin{equation}
W_{\mu\nu} = \sum\limits_n \!\!\int\!\prod_{i=1}^{n}\left[
\frac{d^3p_i}{(2\pi)^32E_i}\right]\,(2\pi)^3\,
\delta^4(p_B-p_e-p_\nu-\sum\limits_i^n p_i)\,\langle\,B\,|\,
j_\nu^\dagger(0)\,|\,n\,\rangle\,\langle\,n\,|\,j_\mu(0)\,|\,B\,
\rangle.
\end{equation}
The summation includes all possible $n-$particle final hadronic
states;
$j_\mu = \bar u\gamma_\mu (1-\gamma_5)b$ is the weak current
mediating
the decay $b\to u$, and  $p_e$, $p_\nu$, $p_B$ are the four-momenta
of
the electron, antineutrino and B-meson, respectively.

The hadronic tensor $W_{\mu\nu}$ can be expressed
 in terms of invariant functions:
\begin{eqnarray}
W_{\mu\nu} & = &
-\!\!\int\frac{d^4x}{2\pi}\,e^{iqx}\,\langle\,B\,| \left[j_\mu(x),
j_\nu^\dagger(0)\right]|\,B\,\rangle\nonumber\\
& = &\frac{1}{\pi}{\rm Im}\,i\!\int\!\! d^4x\,e^{iqx}\,\langle\,B\,
|\,
Tj_\nu^\dagger(0)j_\mu(x)\,|\,B\,\rangle\nonumber\\
& = & {}-g_{\mu\nu} W_1 + p_{B\mu} p_{B\nu} W_2 - i
\epsilon_{\mu\nu\rho\sigma}p_B^\rho q^\sigma W_3 + \dots
\label{eq:Wi}
\end{eqnarray}
The missing terms proportional to different Lorentz structures
do not contribute to the decay rate in the limit of vanishing lepton
mass. The structure functions $W_i$ depend on the invariant mass
squared of the hadrons in the final state, $s = (p_B-p_e-p_\nu)^2$,
and on the invariant mass squared of the lepton pair, $q^2 = (p_e +
p_\nu)^2$.

Integrating over the neutrino momentum, we obtain from
(\ref{eq:gamma}):
\begin{eqnarray}
\Gamma & = & \frac{G_F^2|V_{ub}|^2}{32\pi^3 m_B^2}\!\int\!\!dE_e\,
dq^2\,ds \left\{ q^2 W_1 + m_B^2 \left(\frac{E_e}{m_B}[m_B^2+q^2-s]-
2E_e^2-\frac{q^2}{2}\right)W_2 \right.\nonumber\\
& & \left. + m_B q^2 \left( \frac{m_B^2+q^2-s}{2m_B}-2E_e\right)
W_3\right\}\label{eq:mastereq}
\end{eqnarray}
with the electron energy $E_e$ in the rest-frame of the B-meson.
The
limits of integration are given
by:
\begin{mathletters}
\begin{eqnarray}\label{eq:kinlimits}
0 & \leq & E_e \leq \frac{m_B}{2},\\
0 & \leq & q^2 \leq 2 m_B E_e,\\
0 & \leq & s \leq (m_B - 2 E_e) (m_B - q^2/(2 E_e)).
\end{eqnarray}
\end{mathletters}
The subprocess with the lowest possible value of $s$ is the
production of a single pion in the final state. Hence the structure
functions $W_i$ vanish for $s$ less than $m_\pi^2$ and the
accessible
phase space is restricted to the values depicted in
Fig.~\ref{fig:dalitz}. The shaded area
indicates the region of phase space where $s\leq 1\mbox{\ GeV}^2$,
 the
dashed-dotted lines bound regions with $s \leq 5,10,15,20\,
\mbox{GeV}^2$ (from top to bottom). Since in the present paper the
calculations are carried out in the chiral limit $m_\pi\rightarrow
 0$,
we will in the following stick to the limits of integration given
in
(\ref{eq:kinlimits}).

\section{Derivation of the sum rule}

Our objective is to evaluate the structure functions $W_i$ entering
(\ref{eq:mastereq}) by means of QCD sum rules. To this end we
consider the four-point function
\begin{eqnarray}
T_{\mu\nu} & = & i^3\!\!\int\!\! d^4\!x\,d^4\!y\, d^4\!z e^{-iq_1x -
 i p_1 y + ip_2 z} \langle\, 0\, | \,T j_B(z)j^\dagger_\nu(0)
 j_\mu(x)
j_B^\dagger(y)\,|\,0\, \rangle\nonumber\\
& = & T_g g_{\mu\nu} + T_p \frac{1}{4} (p_1+p_2)_\mu (p_1+p_2)_\nu +
T_\epsilon \frac{1}{2} \epsilon_{\mu\nu\rho\sigma}(p_1+p_2)^\rho
q_1^\sigma + \dots
\label{eq:correlation}
\end{eqnarray}
where $j_B = \bar q i\gamma_5 b$ is the interpolating field of the
B-meson and $j_\mu = \bar u \gamma_\mu (1-\gamma_5) b$ is the weak
current mediating the $b\to u$ decay.

The  four-point function in (\ref{eq:correlation}) can be calculated
in perturbation theory in the deep Euclidean region, i.e.\ for large
negative values of the virtualities
$s$, $(p_1^2-m_b^2)$ and $(p_2^2-m_b^2)$. When
the physical region is approached, nonperturbative effects become
important. The idea underlying the QCD sum rules method is
that at virtualities of $\sim -1\,\mbox{GeV}^2$
the most important nonperturbative corrections can be taken into
account by nonvanishing vacuum expectation values of operators
occurring in the operator product expansion of nonlocal matrix
elements like the correlation function (\ref{eq:correlation}).

In our calculation, we take into account the leading perturbative
contribution, which is given by the box-diagram shown in
Fig.~\ref{fig:contr}(a), and the contributions of the quark and the
mixed condensate, examples of which are given in
Fig.~\ref{fig:contr}(b).

The correlation function (\ref{eq:correlation}) at
$t=(p_1-p_2)^2=0$
and
$q^2 := q_1^2 = (p_2-p_1-q_1)^2$ is related to
hadronic matrix elements via dispersion relations. We express the
invariant
functions $T_i(s,p_1^2,p_2^2,q^2)$ by the dispersion relation
(see e.g.\ \cite{grins})
\begin{equation}
T_i(s,p_1^2,p_2^2,q^2) = \int\limits_{m_\pi^2}^{\infty}\!\!
d\tilde{s}\,
\frac{W_i(\tilde{s},p_1^2,p_2^2,q^2)}{\tilde{s}-s} +
\int\limits_{-\infty}^{s_u}\!\! d\tilde{s}\,
\frac{R_i(\tilde{s},p_1^2,p_2^2,q^2)}{\tilde{s}-s}\, .
\label{eq:x3.2}
\end{equation}
The dispersive part for $s>m_\pi^2$ is due to the insertion of
intermediate states in the direct channel, $\langle\, 0\, |\,
j_B\,j_\nu^\dagger
\, | \, Z_s\,\rangle \, \langle\, Z_s\, |\, j_\mu\, j_B^\dagger\,
 |\,
 0
\,\rangle$, whereas the dispersive part  below $s_u$ comes from the
crossed channel
$\langle\, 0\, |\, j_B\,j_\mu\, | \, Z_u\,\rangle \, \langle\, Z_u\,
 |\,
j_\nu^\dagger\, j_B^\dagger \, |\, 0\,\rangle$. The lowest lying
intermediate state
$Z_u$ contains two heavy quarks und thus
\begin{equation}\label{eq:x3.3}
s_u = p_1^2 +p_2^2 + 2 q^2 - (m_B+m_{B^*})^2\, .
\end{equation}
In the heavy quark limit $m_Q\to\infty ,\: q^2/m_Q^2 < 1$, the
contribution of the crossed channel vanishes, and thus the
structure functions can be related directly to the expansion at
short
distances \cite{totalrate}. On the other hand, for finite quark
masses
 the two cuts may
come very close to each other for on-mass-shell B-mesons and $q^2\to
 m_B^2$. It is specific to the QCD sum rule approach, however,
that calculations are done for sufficiently off-shell values of
$p_i^2$: $p_i^2-m_B^2 \alt -1\mbox{\ GeV}^2$. In this
way the singularities of the direct and the crossed channel are well
separated. Furthermore, the application of this technique requires
$q^2$ to stay several GeV$^2$ below $m_b^2$, the quark mass
squared,
and hence the branching point $s_u$ lies at least 6~GeV$^2$ below
the
the physical region of the B-decay. We shall see later that in our
approach the contribution of the crossed channel can be neglected
(see below Eq.~(\ref{double})). The required information
on the structure functions $W_i(s,q^2)$ of the B-decay is contained
in the terms
with poles in both the variables $p_1^2$ and $p_2^2$ of the
functions
$W_i(s,p_1^2,p_2^2,q^2)$ in Eq.~(\ref{eq:x3.2}):
\begin{equation}\label{eq:x3.3a}
W_i(s,p_1^2,p_2^2,q^2) = \frac{|\langle\, 0\, |\, j_B \, |\, B\,
\rangle|^2
}{(p_1^2-m_B^2)(p_2^2-m_B^2)} \, W_i(s,q^2) + \dots
\end{equation}
Here the dots stand for contributions of higher resonances and the
continuum.

In order to estimate the contribution of higher states in
(\ref{eq:x3.3a}) we use an approximation which is standard in the
QCD
sum rule approach and assume that all contributions of
higher mass states are eliminated by expressing the
four-point correlation function in (\ref{eq:correlation}) by
a double dispersion relation in $p_1^2$ and $p_2^2$ and retaining
only the contribution below a certain threshold $s_0$. The interval
of duality,
i.e.\ the value of the continuum threshold $s_0$, is estimated by
the criterium
of stability of the sum rules.

Following \cite{SVZ}, we subject the correlation function
(\ref{eq:correlation}) to a Borel transformation in both the
variables $p_1^2$ and $p_2^2$.
For an arbitrary function of the Euclidean momentum, $f(P^2)$ with
$P^2=-p^2$, this transformation is defined as
\begin{equation}\label{eq:defborel}
\widehat B\, f = \lim_{\begin{array}{c}\scriptstyle
                  P^2\to\infty,N\to\infty\\[-1mm]
\scriptstyle      P^2/N = M^2\mbox{\scriptsize\ fixed}
                  \end{array}}
            \frac{1}{N!} (-P^2)^{N+1} \frac{d^{N+1}}{(dP^2)^{N+1}}\,
 f
\end{equation}
where $M^2$ is a new variable, called Borel parameter. The objective
 of
this step is to reduce the dependence of the sum rules on the
 unknown
higher-order terms in the operator expansion and on the continuum
model. Indeed, the application of the Borel transformation to a
typical contribution yields
\begin{equation}
\label{eq:y}
\widehat B \,\frac{1}{(p^2-m^2)^n} =
\frac{1}{(n-1)!} (-1)^n \frac{1}{(M^2)^n}
e^{-m^2/M^2},
\end{equation}
and thus contributions of vacuum condensates of high dimension
(which contain high powers of $(p^2-m^2)$ in the denominator) are
suppressed by factorials. In
addition, the contributions of higher mass states become
 exponentially
suppressed, since the Borel transformed expressions for the hadron
propagators $(p^2-m_1^2)^{-1}$ and $(p^2-m_2^2)^{-1}$ differ by an
exponential factor $\exp (-(m_2^2-m_1^2)/M^2)$.
It is important to note that a single Borel transformation in
the variable $p_1^2 =p_2^2 = p^2$ is not sufficient for our
purposes, since it does not suppress terms containing a  pole
in one variable only -- either in $p_1^2$ or in $p_2^2$.

Since, according to Eq.~(\ref{eq:x3.2}), we are further interested
in
comparing $T_{\mu\nu}$ to the dispersion integral in $s$ in terms
of
the structure functions $W_i$ and $R_i$,
we express the amplitudes $T_i$ in form of triple
dispersion integrals in $s$, $p_1^2$ and $p_2^2$. This step requires
some care, since in the region of positive $q^2$ the condition
$t=(p_1-p_2)^2=0$ enforces $p_1^2=p_2^2$. Thus,
we start from negative values of $q^2$, where $p_1^2$ and $p_2^2$
are
independent variables and write
\begin{equation}\label{eq:x}
T_i (p_1^2,p_2^2,s,q^2,t=0) = \int\!\!
d\tilde{s}\,ds_1\,ds_2 \frac{\rho_i(s_1,s_2,\tilde
s,q^2)}{(s_1-p_1^2)(s_2-p_2^2)(\tilde{s}-s)} +\ldots\label{triple}
\end{equation}
As can be shown, (\ref{triple}) defines an analytic function of
$q^2$
and can be continued to positive values of $q^2$. As pointed out in
\cite{BBD}, for positive $q^2$ non-Landau singularities may
 generally
enter the game. These additional contributions are, however,
numerically unimportant in the present analysis and thus have been
omitted, cf.\ the appendix for details.

Due to the condition $t=0$, the
support of the spectral function $\rho_i$ in (\ref{triple}) is
restricted to the line $s_1=s_2$, i.e.\
$\rho_i(s_1,s_2,\tilde{s},q^2) = \rho_i(s_1,\tilde{s},q^2)
\delta(s_1-s_2)$, which means that the four-point function does
not receive any contribution from non-diagonal transitions like
$B\to
B'$. All contributions to the four-point function with a single
pole only in $p_1^2$ or $p_2^2$ cannot be expressed as double
dispersion relation in both $p_1^2$ and $p_2^2$ and
are related to subtraction terms in one of the variables.
All these subtractions are eliminated by the double Borel
transformation.

Putting together (\ref{eq:correlation}),
(\ref{eq:x3.2}) and (\ref{eq:x}) we get, after performing the Borel
transformation:
\begin{equation}
|\langle\,0\,|\,j_B\,|\,B\,\rangle|^2\,e^{-2m_b^2/M^2}\left\{
\int\limits_{m_\pi^2}^{\infty}\!\!d\tilde{s}
\frac{W_i(\tilde{s},q^2)}{\tilde{s}-s} +
\int\limits_{-\infty}^{s_u}\!\!d\tilde{s}
\frac{R_i(\tilde{s},q^2)}{\tilde{s}-s}
\right\}= \int\!\!
d\tilde{s}\!\!\int\limits_0^{s_0}\!\! ds_1\frac{\rho_i(s_1,\tilde
s,q^2)}{\tilde{s}-s}\,e^{-2s_1/M^2}\label{double}
\end{equation}
where $R_i(s,q^2)$ is defined analogously to $W_i(s,q^2)$ in
(\ref{eq:x3.3a}).

Staying strictly in the framework of the operator product expansion
we cannot
discriminate distributions from the direct and the crossed channel.
We, therefore, have to invoke a concept of duality and assume that
the
relation (\ref{double}) holds true for the spectral densities
themselves, if the latter are smeared over a sufficiently large
interval of $s$. The leading contributions to
the operator
product expansion of (\ref{eq:x}) do not contribute to the
discontinuities
$R_i$ of the crossed channel; for this reason the crossed channel
singularities
do not occur in our calculations.

Existing calculations of deep inelastic structure functions by
QCD sum rules \cite{ioffe,kh} use the approximation of local
 duality,
equating in the expressions analogous to (\ref{double})
the structure functions and the spectral densities
pointwise in $s$. Since we are interested in the electron spectrum
 and
integrate over the neutrino momentum, we automatically are led to an
integration over $s$, cf.\ Eq.~(\ref{eq:mastereq}). Therefore, in
 this
paper we only make the much weaker assumption, that the physical
spectral density coincides to the one obtained by the operator
 product
expansion upon the integration over a sufficiently large interval
of
duality $\Delta s$ as mentioned before. Since the lowest mass
physical
state is the pion,
we demand $s_{max}$ to be larger than the interval of duality in the
QCD sum rules for the pion $\Delta s \simeq 1\,\mbox{GeV}^2$
\cite{SVZ}. Since at $E_e\approx E_{max}$ the physical region in $s$
shrinks to a point, the requirement of having a sufficient interval
 of
duality excludes a certain fraction of the kinematical region, which
we show as a gray area in the Dalitz-plot Fig.~\ref{fig:dalitz}.
For
small values of $E_e$ it covers only a small fraction of the
integration region in $q^2$ and is heavily suppressed by phase
space,
so that this
limitation is not restrictive. More important constraints on the
accessible values of $q^2$ follow for higher electron energies.
However, the major part of the excluded region in the Dalitz plot
is not accessible to the sum rules for a different reason as well:
the value of $q^2-m_b^2$ should be sufficiently large and negative
in
order the operator product expansion to be justified. Taken
 altogether,
these requirements lead to an upper bound for accessible electron
energies of $2.1\,\mbox{GeV}$. At this energy requiring an interval
 of
duality larger than $\Delta s \sim 1\,\mbox{GeV}$ excludes 4.7\% of
the whole phase space in $s$ and
$q^2$. At lower electron energies it is an even smaller fraction.

Expanding (\ref{double}) in powers of energy one may relate the
moments of
structure functions, $\int ds \, s^{-n} W_i(s)$, to
corresponding integrals over the spectral densities, and, in turn,
 to
matrix elements of local operators over the B-meson, which appear in
the short-distance expansion of the $T$-product of weak currents.
The operator expansion of this product is the starting point in
approaches to the inclusive B-decays based on the $1/m_b$ expansion
\cite{totalrate}, or the parton model \cite{bareiss}. It is
worthwile to note that the integration
over $s$ in the differential decay rate at fixed $E_e$ in
(\ref{eq:mastereq}) is constrained by the value
\begin{equation}
s_{max} =(m_B-2E_e)\left(m_B-\frac{q^2}{2E_e}\right)
\end{equation}
which for not too large $E_e$ is less than the actual support of
the
structure functions. Thus, the knowledge of the first few moments
of the structure functions $W_i$ is not sufficient
for the calculation of the inclusive spectrum, and we cannot reduce
our task to the calculation of three-point instead of
four-point functions.

Writing the lepton decay constant $\langle\, 0\, |\, j_B \, |\, B\,
\rangle$ as $m_B^2 f_B/m_b$, we get from (\ref{double}) the
 following
sum rules for the structure functions $W_i$:
\begin{equation}\label{eq:sr}
W_i(s,q^2) =\frac{m_b^2}{m_B^4 f_B^2}
\int\limits_{0}^{s_0}\!\! ds_1 \, \rho_i(s_1,q^2,s)\,
e^{-2(s_1-m_B^2)/M^2}.
\end{equation}
These equations should be understood in the sense that the physical
$W_i$ coincide with the ones given in (\ref{eq:sr}) after smearing
over an interval of duality in $s$ larger or of the order of
1 GeV$^2$.

The spectral densities $\rho_i$ receive contributions from
perturbation theory and the vacuum condensates
\begin{equation}
\rho_i =
\rho_i^{pert} + \rho_i^{\langle 3 \rangle} \langle\,\bar q q \,
\rangle + \rho_i^{\langle 5
\rangle} \langle\,\bar q \sigma g G q \,\rangle.
\end{equation}
A calculation of the box graph in Fig.~\ref{fig:contr}(a) yields
(see appendix for the details)
\begin{mathletters}
\begin{eqnarray}
\rho_1^{pert} & = & {}-\frac{3}{4\pi^2}
\frac{m_b^2-s_1}{q^2}\left( m_b^2 + \frac{2 q^2 s +(m_b^2-q^2)
( s + q^2-s_1)}{\lambda^{1/2}}\right)\theta(s_1-s_1^L),\\
\rho_2^{pert} & = & \frac{3}{\pi^2}\, (m_b^2-s_1)\,\frac{2 q^2 s +
(m_b^2-q^2) ( s +q^2-s_1)}{\lambda^{3/2}}\theta(s_1-s_1^L),\\
\rho_3^{pert} & = &\frac{3}{2\pi^2} \frac{m_b^2-s_1}{\lambda^{1/2}}
\theta(s_1-s_1^L),
\end{eqnarray}
\end{mathletters}
where
\begin{equation}\label{eq:?}
s_1^L = m_b^2 + \frac{s m_b^2}{m_b^2-q^2}.
\end{equation}
Note that $\rho_1^{pert}$ is regular in the limit $q^2\to 0$.

The contribution of the quark condensate is given by
\begin{mathletters}
\begin{eqnarray}
\rho_1^{\langle 3 \rangle} & = & {}-2 \, m_b(s + m_b^2 - q^2)\,
\delta(s_1-m_b^2)\,\delta(s),\\
\rho_2^{\langle 3 \rangle} & = &  -8 m_b \,\delta(s_1-m_b^2)\,
\delta(s),\\
\rho_3^{\langle 3 \rangle} & = & 4 m_b \, \delta(s_1-m_b^2)\,
\delta(s),
\end{eqnarray}
\end{mathletters}
and the contribution of the mixed condensate reads
\begin{mathletters}
\begin{eqnarray}
\rho_1^{\langle 5 \rangle} & = & m_b \left\{
-\left[\frac{1}{3}\delta(s)+(m_b^2-q^2)\delta'(s) \right]
\delta(s_1-m_b^2)-\left[ 2(m_b^2-q^2)\delta(s)
\right.\right.\nonumber\\
 &&\mbox{}-\left.\left.
\frac{1}{2}(m_b^2-q^2)^2\delta'(s) \right] \delta'(s_1-m_b^2)
+\frac{m_b^2(m_b^2-q^2)}{2}\delta(s)\delta''(s_1-m_b^2)\right\},\\
\rho_2^{\langle 5 \rangle} & = & m_b \left\{
-\frac{8}{3}\delta'(s)\, \delta(s_1-m_b^2)
+ \left[-4\delta(s)+2(m_b^2-q^2)\delta'(s) \right]
\delta'(s_1-m_b^2)
\right.\nonumber\\
 &&\left. {}+2m_b^2\delta(s)\, \delta''(s_1-m_b^2)\right\},\\
\rho_3^{\langle 5 \rangle} & = & m_b \left\{\delta'(s)\,
\delta(s_1-m_b^2) + \left[3\delta(s) - (m_b^2-q^2) \delta'(s)\right]
\,
\delta'(s_1-m_b^2)\right. \nonumber\\
& & \left. {} - m_b^2 \delta(s) \, \delta''(s_1-m_b^2)\right\}.
\end{eqnarray}
\end{mathletters}

Note that $\delta(s)$ has to be understood as the limit of
$\delta(s-m_u^2)$ where we have put $m_u$, the mass of the u-quark,
 to
zero. Thus there is no ambiguity in the integration over $s$ within
the range specified by (\ref{eq:kinlimits}). Contributions of the
 u-quark
condensate $\langle \bar u u\rangle$ and of the corresponding mixed
condensate $\langle \bar u\sigma g G u \rangle$
vanish. Note that the contributions containing $\delta'(s)$ give
rise
to terms $\sim\delta(2E_e(m_B-m_u^2/(m_B-2E_e)) - q^2)$ in the
 double
differential rate $d^2\Gamma/(dE_edq^2)$.

The leptonic decay constant $f_B$ entering Eq.~(\ref{eq:sr}) is
determined by a two-point sum rule \cite{alel} where the same
approximations are made and the same input parameters are used as
for
the sum rule (\ref{eq:sr}), except for the value of the
 Borel parameter $M_{2pt}^2$, which is taken
 two times smaller, $M_{2pt}^2 = M^2/2$ \cite{BBD}:
\begin{eqnarray}
\lefteqn{f_B^2 m_B^4\, e^{(m_b^2-m_B^2)/M^2_{2pt}} =}\label{eq:fB}\\
& = & \frac{3m_b^2}{8\pi^2} \int\limits_{m_b^2}^{s_0}\!\! ds_1
\frac{(s_1-m_b^2)^2}{s_1} \,e^{(m_b^2-s_1)/M^2_{2pt}} - m_b^3\,
\langle\,\bar
q q \,\rangle + \frac{m_b^3}{2} \left(\frac{m_b^2}{2M^4_{2pt}} -
\frac{1}{M^2_{2pt}}\right) \langle\,\bar q\sigma g G q \,\rangle.
\nonumber
\end{eqnarray}
Considering the ratio of sum rules instead of fixing $f_B$ at a
certain value considerably reduces the dependence of the results
on the input value of the b-quark mass and on the value of the
Borel parameter, since the effect of a change in these parameters
is
cancelled between the numerator and the denominator to a great
 extent.

Radiative corrections to the sum rule in (\ref{eq:sr}) may in
general
be important, cf. \cite{B^3D,B^2G}. In the limit of large $m_b$,
however, all radiative corrections factorize into corrections to
$f_B$, which are cancelled by taking the ratio of the four-point
and two-point sum rule, and Sudakov-type radiative corrections to
the spectrum of the decay of a free b-quark, which we take into
 account
as a multiplicative factor, see next section. A full account
for the radiative correction requires a laborous calculation of the
$\alpha_s$-correction to the box graph and is beyond the tasks of
 this
paper. On formal grounds, all corrections which we do not take into
account are suppressed by a power of the heavy quark mass.

\section{The heavy quark limit}

We consider the region where the invariant mass of
the lepton pair $q^2$ and the electron energy
$2E_e$ constitute a finite fraction of the heavy
quark mass, i.e.
\begin{eqnarray}\label{eq:scale}
 m_b^2 - q^2& =& {\cal O}(m_b^2),
\nonumber \\
  m_b - 2E_e &=& {\cal O}(m_b).
\end{eqnarray}
It is convenient to introduce the scaling variable \cite{accm}
\begin{equation}
x=2E_e/m_B
\end{equation}
in terms of which the last of the conditions in (\ref{eq:scale})
reads $ 1-x = {\cal O}(1)$.
In this kinematical range the sum rules (\ref{eq:sr})
simplify drastically in the limit of infinite heavy quark mass
and yield the results of the free quark decay.

To show this, we use the common assumption that all excitation
energies remain finite in the heavy quark limit, and hence $s_0$,
the continuum threshold, scales as
\begin{equation}\label{eq:thres}
s_0 - m_b^2 = (m_B+E_{exc})^2-m_b^2 \simeq {\cal O}(m_b).
\end{equation}
The working region in the Borel parameter is determined by the
requirement that the exponential suppression factor for the
contributions of excited states, $\exp[-(s_0-m_B^2)/M^2]$,
remains finite in that limit.
Therefore, also the Borel parameter $M^2$ should be taken at values
 of
order ${\cal O}(m_b)$. Next, we note that the support of the
 structure
functions is restricted to an interval $0 < s < {\cal O}(m_b)$.
As far as the perturbative contribution to the sum rules is
 concerned,
this statement follows from the kinematical restriction for
the double discontinuity of the box graph, $s_1 >s_L$, in
(\ref{eq:?}), where from
\begin{equation}\label{eq:condi}
s \leq (s_1 - m_b^2) \frac{m_b^2-q^2}{m_b^2} \leq {\cal O}(m_b).
\end{equation}
The last inequality is a consequence of Eq.~(\ref{eq:thres}).
Therefore $s$ can be neglected against $s_1$,
which is of ${\cal O}(m_b^2)$.
The nonperturbative contributions generally form a
series in $\delta$-functions of $s$ and its derivatives, which
should
in principle produce a certain smooth function upon summation.

Retaining the leading terms in the heavy-quark mass only, we find
{} from
(\ref{eq:sr}):
\begin{eqnarray}
\frac{f_B^2 m_B^4}{m_b^2} W_1(s,q^2) &=&
{}\frac{3}{4\pi^2} \int_{s_1^L}^{s_0} ds_1\, e^{2(m_B^2-s_1)/M^2}
(s_1-m_b^2)\nonumber\\
&& \mbox{}-
2\langle\,\bar q q \,\rangle m_b (m_b^2-q^2) e^{2(m_B^2-m_b^2)/M^2}
 \delta(s)
\nonumber\\ && \mbox{}+
2\langle\,\bar q \sigma g G q \,\rangle \frac{m_b^3}{M^4}
 (m_b^2-q^2) e^{2(m_B^2-m_b^2)/M^2} \delta(s)\label{eq:hql}
\end{eqnarray}
In order to obtain the inclusive electron spectrum, we integrate
(\ref{eq:hql}) over $s$. Changing the order of integration
in $s$ and in $s_1$, we find the integration intervals
\begin{eqnarray}
m_b^2 &\leq& s_1 \leq s_0,
\nonumber\\
\qquad 0&\leq& s \leq min\left(
\frac{m_b^2-q^2}{m_b^2} (s_1-m_b^2),\: (m_b-2E_e)(m_b -
\frac{q^2}{2E_e})\right).
\end{eqnarray}
Due to (\ref{eq:condi}), the second term in $min(\ldots,\ldots)$ is
parametrically large as compared to the first one. Hence the full
support of $W_1$ in $s$ is covered, exemplifying the general
 statement
that in the heavy quark limit the knowledge of the first moment
$\int ds\,W_1(s)$ is sufficient for the calculation of the electron
spectrum in the kinematical range given in (\ref{eq:scale}), cf.\
Refs.~\cite{totalrate}. Thus, we obtain:
\begin{eqnarray}
\frac{f_B^2 m_B^4}{m_b^2} \int ds \,W_1(s,q^2) &=&
2(m_b^2-q^2)\left\{
\frac{3}{8\pi^2} \int_{m_b^2}^{s_0} ds_1\, e^{2(m_B^2-s_1)/M^2}
\frac{(s_1-m_b^2)^2}{m_b^2}
\right.\nonumber\\ && \mbox{}-\left.
\langle\,\bar q q \,\rangle m_b e^{2(m_B^2-m_b^2)/M^2}
+\langle\,\bar q \sigma g G q \,\rangle \frac{m_b^3}{M^4}
 e^{2(m_B^2-m_b^2)/M^2} \right\}. \label{eq:hql1}
\end{eqnarray}
Letting $M^2 = 2 M_{2pt}^2$, one realizes that the expression in
 curly
brackets coincides (apart from an overall factor $m_b^2$) with the
leading term of the two-point sum rule (\ref{eq:fB}) in the limit
of
infinite quark mass. The same factorization takes place in the other
two structure functions. Thus, in the heavy quark limit we end up
 with
the following expressions:
\begin{mathletters}
\begin{eqnarray}
 \int ds \,W_1(s,q^2) &=& 2(m_b^2-q^2),\\
 \int ds \,W_2(s,q^2) &=& 8,\\
 \int ds \,W_3 (s,q^2) &=& 4,
 \end{eqnarray}
\end{mathletters}
which coincide with the structure functions for
the decay of a free heavy quark.
The corresponding electron spectrum is \cite{accm}
\begin{equation}\label{eq:uncorr}
\frac{d\Gamma}{dx}=
\frac{ G^2 |V_{ub}|^2 m_b^5}{96\pi^3} x^2(3-2x),
\end{equation}
where $x=2E_e/m_b$, and the total decay rate equals
\begin{equation}
\Gamma=
\frac{ G^2 |V_{ub}|^2 m_b^5}{192\pi^3}.
\end{equation}
Note that to this accuracy
there is no distinction between the
quark and the meson mass.

The results presented above do not take into account radiative
corrections $\propto \alpha_s(m_b)$. As it is well known, such
corrections can in general get enhanced by large logarithms of the
heavy quark mass. The corresponding corrections to the spectrum
for the decay of a free quark have been calculated in \cite{accm}.
They result in a modification of the uncorrected
spectrum (\ref{eq:uncorr}) by a multiplicative factor:
\begin{equation}\label{eq:radcorr}
\frac{d\Gamma}{dx}= \frac{d\Gamma^{(0)}}{dx}
\left [1-\frac{2\alpha_s}{3\pi}\widetilde{G}(x)\right]
\exp\left [ -\frac{2\alpha_s}{3\pi}\ln^2 (1-x)\right],
\label{eq:factor}
\end{equation}
where the Sudakov-like exponential has been factored out and the
function $\widetilde{G}$ is given by
\begin{eqnarray}
\widetilde{G}(x) & = & \frac{1}{x} \left\{ \ln (1-x) \left[
\frac{3}{2} \ln
\frac{1+2x}{3} + \frac{11}{6x} + \frac{1+4x}{3}\right] + 2 x \,{\rm
 Li}\,
(x)\right.\nonumber\\
& & \left. {}+ \frac{3}{2}\, {\rm Li}\left[ \frac{2}{3} (1-x)
 \right] -
\frac{3}{2}\, {\rm Li}\left( \frac{2}{3} \right) + x \left(
 \frac{2}{3}
\,\pi^2 - \frac{19}{12} \right) + \frac{11}{6} \right\}.
\label{eq:G}
\end{eqnarray}
In numerical calculations, the results of which are presented below,
 we
take into account this Sudakov-type correction, which is the leading
one in the kinematical region in (\ref{eq:scale}). All the radiative
corrections due to the spectator quark are suppressed by a power of
$m_b$. It should be noted, however, that for the region of high
 electron
energy $1-x \sim 1/m_b$ the application of (\ref{eq:radcorr}) is
not justified, and the true effect of radiative corrections may be
different, as indicated by the qualitative
change of the Sudakov-type exponential
behaviour with the account of the constituent mass of the u-quark
\cite{accm}.

\section{Results and Discussion}

The results of the numerical calculations given below have been
 obtained
using the following
values of the vacuum condensates:
\begin{eqnarray}\label{eq:condensates}
\langle\,\bar q q \,\rangle (1\,\mbox{GeV}) = (-0.24\,\mbox{GeV})^3,
\nonumber\\
\langle\,\bar q \sigma g G q \,\rangle (1\,\mbox{GeV})=
0.8\,\mbox{GeV}^2\,\langle\,\bar q q \,\rangle (1\,\mbox{GeV}).
\end{eqnarray}
These values have been
rescaled to a higher normalization point $\mu^2=M^2$,
which is the relevant short distance expansion parameter
in the sum rule (\ref{eq:sr}). We have varied the value of the pole
mass of the b-quark $m_b$ within the range
(4.6 -- 4.8)~GeV, which covers the values given in literature
\cite{voloshin,reinders}.
 The continuum threshold $s_0$ is
determined from the requirement of stability of
 the two-point sum rule in (\ref{eq:fB}) and depends on
the quark mass. We use the values
$s_0(m_b = 4.6\mbox{\ GeV}) = 36 \mbox{\
GeV}^2$, $s_0(4.7\mbox{\ GeV}) = 35 \mbox{\ GeV}^2$ and
$s_0(4.8\mbox{\ GeV}) = 34 \mbox{\ GeV}^2$ \cite{B^3D}. These values
 provide a good stability of the two-point sum rule in a large
range of Borel parameters
$M_{2pt}^2$. Note that we use the same value of
$s_0$ in both the
four-point and the two-point sum rules.
 Similarly, the range of appropriate values
of the Borel parameter $M^2$ in the four-point sum rule is fixed
{}from the two-point sum rule (\ref{eq:fB}) by the requirement that
 both
the contribution of the continuum and of nonperturbative
corrections do not
exceed 40\% each. In this way we come to the range
$3.5\mbox{\ GeV}^2 \alt M^2_{2pt}
\alt 5\mbox{\ GeV}^2$. Similar to the case of the three-point
function \cite{BBD},
 the Borel parameter of the four-point
function, $M^2$, should be taken two times larger than
 $M^2_{2pt}$, i.e. we arrive at the interval of values
$7\mbox{\ GeV}^2 \alt M^2 \alt 10\mbox{\ GeV}^2$.

The QCD sum rule predictions for the differential spectrum
$d\Gamma/dx$ in the rest frame of the B-meson are presented
in Fig.~\ref{fig:spec.M} in dependence on the scaling  variable
$x = 2E_e/m_B$ for different values of the Borel parameter, and for
input values $m_b = 4.8\mbox{\ GeV}$ and $s_0 =
34\mbox{\ GeV}$.
The
curves are calculated by means of Eq.~(\ref{eq:mastereq}) up to
the electron energy $2.1\,$GeV ($x=0.80$),
where the upper limit of
integration in $q^2$ is still small enough as to ensure the
validity of the operator product expansion (cf.\ Sect.~III). For
higher electron energies we did a smooth interpolation to the end
 point
of the spectrum at $x=1$,
where $d\Gamma/dx=0$ owing to vanishing phase space.
The position of the maximum of the spectrum is obtained
at values $x\sim 0.75$, which are still
 within the region of validity of the QCD sum rule
calculation, and due to that the
uncertainty introduced by the degree of the interpolating
polynomials
is negligible: it turns out to be
smaller than 1\% in the experimentally interesting region $2.2
\mbox{\ GeV}\leq E_e\leq 2.4\mbox{\ GeV}$ when changing from a
cubic
to an order five
polynomial. We find that the variation of the Borel parameter
within the range given above induces a
$\pm 15\%$ uncertainty in the absolute normalization, while the
shape
of the spectrum is much more stable.

Fig.~\ref{fig:spec.cont} shows the separate contributions to the
differential spectrum coming from the perturbative graph and from
the
nonperturbative corrections. The contributions of the vacuum
condensates are large by themselves, but partly cancel in the sum.
Up
to the very end of the spectrum the relative weights of the
perturbative and nonperturbative contributions remain practically
constant.

A variation of the mixed condensate in the range $0.6\,\mbox{GeV}^2
\leq m_0^2 \leq 1\,\mbox{GeV}^2$ leads to a change of the decay rate
by $\pm 10\%$.

The influence of the mass of the heavy quark, which is
large in the
case of the sum rule for $f_B$ \cite{B^3D,narr}, is eliminated to
a great extent
by compensations between
 the numerator and the denominator in (\ref{eq:sr}). A
variation of $m_b$ from 4.8 to 4.6$\,$GeV reduces the decay rate by
8\%. A similar compensation occurs for the dependence on the
 continuum
threshold $s_0$: variation between 30 and 38$\,\mbox{GeV}^2$ results
in a change by $\pm4\%$. Variations of other parameters within
the standard limits have only marginal effect.

The radiative corrections in (\ref{eq:factor}) reduce the decay rate
 by
20\%. Since the main effect is however achieved at large values of
 $x$, where neither the
derivation of (\ref{eq:factor}) nor the QCD sum rules approach are
justified, we add an additional error of 10\% to our results, which
corresponds to an uncertainty in the radiative corrections of 50\%.

It is worthwhile to mention that major part of uncertainties of the
QCD sum rule calculation result in an uncertainty in the
overall normalization factor,
but not affect the shape of the spectrum.

Combining all sorts of errors, we obtain the total decay rate:
\begin{eqnarray}
\Gamma &=& (4.5\pm 1.3)\, 10^{-11} |V_{ub}|^2\,\mbox{GeV},
\nonumber\\
& = & (6.8\pm 2.0)\, 10^{13}\,|V_{ub}|^2\,\mbox{s}^{-1}.
\end{eqnarray}
In order to be able to compare to the experimental results which
were
obtained for B-mesons created on the $\Upsilon(4S)$ resonance, we
have to
boost to the laboratory frame,
\begin{equation}
\frac{d\Gamma^{lab}}{dE_e} = \int\limits_{-v_B E_e}^{+ v_B E_e}\!\!
du
\,\frac{1}{2v_B E_e}\,\frac{d\Gamma^{rest}}{dE_e}(E_e+u)\, ,
\end{equation}
where the electron energy $E_e$ is measured in the laboratory frame
and
$v_B = 0.065$ is the velocity of the B-meson. The change induced by
the boost
is negligible for the total rate, but modifies
the high energy part of the spectrum. Thus,
in Table~I we give the values of the integrated spectrum in energy
bins of
0.1~GeV from 2.1~GeV to 2.7~GeV both in the rest and the lab-frame.

For the integrated value of the decay rate in the interval of
electron energies $2.4\, \mbox{GeV} \leq E_e \leq 2.6\,\mbox{GeV}$
we
 get:
\begin{equation}
\Gamma^{lab} = (5.0\pm 1.6)\,10^{12}\, |V_{ub}|^2\, \mbox{s}^{-1}.
\end{equation}
Combining the latter value with the new experimental result of CLEO
\cite{cleo}
for the branching ratio in the same region of energies,
\begin{equation}
B  = (0.53\pm 0.14\pm 0.13)\, 10^{-4},
\end{equation}
 we obtain the following value for the CKM-matrix element:
\begin{equation}
|V_{ub}| = (0.003\pm 0.001)
\left(\frac{1.3\,\mbox{ps}}{\tau_B}\right)^{1/2}\, ,
\end{equation}
where $\tau_B$ is the lifetime of the B-meson. The error
takes into account the accuracy in the calculated decay rate,
a conservative estimate for the uncertainty induced by the
extrapolation and the experimental error, where theoretical and
experimental error are roughly of the same size.

The comparison with the other existing calculations of absolute
rates
is given in Table~I, and the comparison of the predicted shape of
the
spectrum is shown in Fig.~\ref{fig:radcorr}. The QCD sum rule
calculation of the electron spectrum for $2.1\,\mbox{GeV} <
E_e^{lab} <
2.5\,\mbox{GeV}$ lies inbetween the predictions of the free
\cite{accm} and the modified \cite{accm,RDB} and the direct
summation
of the contributions of exclusive channels \cite{GISW}, while for
higher energies our results coincide with \cite{accm} and
\cite{RDB}
within the errors. At such high
energies, however, the extrapolation procedure used might not be
adequate. All calculations of the shape of the spectrum are in a
good
agreement at low values of $x$, while starting from $x\sim 0.5$ the
discrepancies amount (10--15)\%. The calculations  using a certain
modification of the free quark model have an endpoint which might
differ from the physical one, since the quark masses enter in this
approach instead of the hadron ones. The slope of all curves near
the
end is, however, not so much different, the one of \cite{GISW}
being
the smallest.

Concluding we remark that the sum rule method allows a
determination of the semileptonic decay rate which is on a
theoretically sound basis for electron energies below
$2.1\,\mbox{GeV}$. Though at first sight it may seem unsatisfactory
that the most interesting range of electron energies can
be reached only by a constrained extrapolation, this does not induce
a major error. This is due to the fact, that the electron spectrum
 can
be calculated to energies larger than the position of the  maximum
 and
that the endpoint of the spectrum, where it vanishes, is determined
 by
hadron, and not quark masses in our approach. In the determination
of the CKM matrix element $V_{bu}$ the uncertainty of the lifetime
 is
still the most significant error.

\acknowledgments

We thank Franz Muheim for providing us with the new CLEO data prior
to publication.
One of us (P.B.) gratefully acknowledges financial support by the
Deutsche Forschungsgemeinschaft (DFG).

\unletteredappendix{The Triple Dispersion Relation of the
 box-diagram}

In order to be able to subtract the contribution of the continuum
{} from
our sum  rules and to take the imaginary part in $s$, the invariant
mass squared of the final state hadrons, we need to represent the
box-diagram  Fig.~\ref{fig:contr}(a) by a triple dispersion
relation.
Since the forward scattering amplitude is relevant to our
problem, we consider the kinematical configuration $t =
(p_1-p_2)^2 = (q_1-q_2)^2 \equiv 0$. In addition to that, we want to
keep $p_1^2 \neq p_2^2$ whilst $q_1^2 \equiv q_2^2 \equiv q^2$ which
implies $q^2<0$. Thus, in our case the independent variables
determining the box-diagram are $p_1^2$, $p_2^2$, $q^2$ and $s$. The
external momenta of the interpolating field of the
B-meson, $p_1$ and $p_2$, are considered in the Euclidean region
$p_1^2,p_2^2 < 0$.

We infer the existence of such a triple dispersion relation for the
fermionic case from the analysis of the scalar box-diagram for
 $q^2<0$
\begin{eqnarray}
S & = & \int\!\!
\frac{d^4k}{(2\pi)^4} \frac{1}{((p_1+k)^2-m_b^2) ((p_2+k)^2-m_b^2)
(p_1+q_1)^2 k^2} \nonumber\\
& = & \frac{i}{16\pi^2}\frac{1}{m_b^2-q^2}\int\!\! d\tilde s
\frac{1}{\tilde s -s} \int\!\! ds_1 \int\!\! ds_2
\frac{\delta(s_1-s_2)}{(s_1-p_1^2)(s_1-p_2^2)}\nonumber\\
& = & \frac{i}{16\pi^2}\frac{1}{m_b^2-q^2} \int\limits_0^\infty\!\!
 d\tilde s
\frac{1}{\tilde s -s} \int\limits_{s_1^L}^\infty \!\! ds_1
\frac{1}{(s_1-p_1^2)(s_1-p_2^2)}\label{eq:disscalar}
\end{eqnarray}
with
\begin{equation}
s_1^L = \frac{\tilde s m_b^2}{m_b^2-q^2} + m_b^2.
\end{equation}
The right hand side of (\ref{eq:disscalar}) can be continued
analytically to $q^2>0$ with $p_1^2$ and $p_2^2$ remaining
 independent
variables. We have checked this issue also explicitly by comparing
with the corresponding Feynman-parameter integral. For arbitrary
values of $q^2$, Eq.~(\ref{eq:disscalar}) can equally well be
 obtained
{}from the scalar triangle graph $T$ with external momenta squared
$p_1^2$, $q^2$ and $s$ as
\begin{equation}
S = \left.\frac{d}{dm_b^2} T(p_1^2,s,q^2)\right|_{(s_1-p_1^2)^2\to
 (s_1-p_1^2)
(s_1-p_2^2)}\label{eq:derivative}
\end{equation}
which follows immediately from the expressions for the
 loop-integrals.

As pointed out in \cite{BBD}, a na\"{\i}ve continuation of the
double
dispersion relation for the triangle graph to the region of positive
values of $q^2$ is not possible because of the existence of
 non-Landau
singularities. Technically, the problem arises because the double
spectral function of the triangle graph $T$ with respect to $p_1^2$
 and $s$,
$\rho_T(p_1^2,q^2,s)$, contains a factor $1/\lambda^{1/2}$, where
 $\lambda
= p_1^4 + s^2 + q^4 - 2 p_1^2 s - 2 p_1^2 q^2 - 2 s q^2$. For
 positive
values of $q^2$, $\lambda$ can vanish and $\rho_T$ can become
 singular
at the boundary of the integration region as given by the Landau
equations. This allows the
logarithmic branching cut on the unphysical sheet of the square root
to dive through the square root cut into the physical sheet. The
proper treatment of that additional singularity is described in
 detail
in \cite{BBD}. As follows from Eq.~(\ref{eq:disscalar}), for the
scalar box-graph the non-Landau singularities are absent, but are
present for fermions. They play, however, no role in our
analysis, because it turns out that for physically relevant $q^2$
 they
lie above the continuum threshold or are numerically negligible due
to phase space suppression.

It proves convenient to apply the above method of taking the
 derivative
of some triangle diagram with appropriate vertices with respect to
$m_b^2$ likewise to the fermionic case with open Lorentz-indices.
{} From
the double spectral functions for the tensor-integrals given in
\cite{thesis} we obtain (for $q^2<0$)
\begin{eqnarray}
\lefteqn{\int\!\!\frac{d^4k}{(2\pi)^4}
\frac{k_\mu}{((p_1+k)^2-m_b^2)
((p_2+k)^2-m_b^2) (p_1+q_1)^2 k^2} =}\nonumber\\
& = &  P_+ \frac{1}{2}(p_1+p_2)_\mu + P_- \frac{1}{2}(p_1-p_2)_\mu +
Q_1 q_{1\mu}.
\end{eqnarray}
The only invariant needed is $P_+$ for which we obtain:
\begin{equation}
P_+ = -S + \frac{i}{16\pi^2} \int\limits_0^\infty\!\! d\tilde s
\frac{1}{\tilde s -s} \int\limits_{s_1^L}^\infty \!\! ds_1
\frac{1}{\lambda^{1/2}(s_1-p_1^2)(s_1-p_2^2)}
\end{equation}
with $\lambda = s_1^2 + \tilde{s}^2 + q^4 - 2 s_1 \tilde s - 2
s_1 q^2 - 2 \tilde s q^2$.

The integral with two open Lorentz-indices is decomposed as
\begin{eqnarray}
\int\!\!\frac{d^4k}{(2\pi)^4} \frac{k_\mu k_\nu}{((p_1+k)^2-m_b^2)
((p_2+k)^2-m_b^2) (p_1+q_1)^2 k^2}=  \nonumber\\
= A g_{\mu\nu} + B \frac{1}{4} (p_1+p_2)_\mu (p_1+p_2)_\nu + \dots
\end{eqnarray}
with
\begin{mathletters}
\begin{eqnarray}
A & = & -\frac{i}{16\pi^2} \frac{1}{4q^2} \int\limits_0^\infty\!\!
 d\tilde s
\frac{1}{\tilde s -s} \int\limits_{s_1^L}^\infty \!\! ds_1
\frac{(m_b^2-q^2)\lambda^{1/2} + 2 q^2 \tilde s + (m_b^2-q^2)
 (\tilde s +
q^2 - s_1)}{\lambda^{1/2}(s_1-p_1^2)(s_1-p_2^2)}\makebox[1cm]{}\\
B & = & -\frac{i}{16\pi^2} \int\limits_0^\infty\!\! d\tilde s
\frac{1}{\tilde s -s} \int\limits_{s_1^L}^\infty \!\! ds_1
\frac{1}{(s_1-p_1^2)(s_1-p_2^2)}\nonumber\\
& & \times\left\{ -\frac{1}{m_b^2-q^2} + \frac{2}{\lambda^{1/2}}
 + \frac{(m_b^2-q^2)(q^2+\tilde s -s_1) + 2 \tilde s
 q^2}{\lambda^{3/2}}
\right\}
\end{eqnarray}
\end{mathletters}
which are the only invariants needed.

The expressions given above (including non-Landau singularities for
$q^2>0$) have been checked for arbitrary values of $q^2$ by
 comparison
with the corresponding Feynman-parameter integrals and by an
independent calculation of the double spectral densities using the
decomposition of momenta in light-cone variables.

By means of the above expressions we obtain the perturbative double
spectral functions given in Sec.~III.

\figure{Dalitz-plot of the phase space in the plane of invariant
lepton mass squared, $q^2$, and electron energy, $E_e$. The solid
border-line where $s = m_\pi^2$ can
only be reached with a single pion in the final hadronic state. The
shaded area shows the region of phase space with $s\leq 1\mbox{\
GeV}^2$. The dashed-dotted lines denote curves of constant $s \geq
5,10,15,20\mbox{\ GeV}^2$ (from top to bottom).\label{fig:dalitz}}
\figure{Feynman diagrams contributing to the operator product
expansion of (\ref{eq:correlation}): (a) perturbative contribution,
(b) contribution of the quark and part of the contribution of the
mixed condensate. Lines ending in a cross
denote vacuum expectation values.\label{fig:contr}}
\figure{The spectrum $|V_{ub}|^{-2} d\Gamma/dx$ as function of $x$
 with
$m_b = 4.8\mbox{\ GeV}$ and $s_0 = 34\mbox{\ GeV}^2$ for different
 values
of the Borel parameter, $M^2 = 7,8,9,10\mbox{\ GeV}^2$ (from top to
bottom). Solid lines: evaluation using (\ref{eq:sr}), dashed lines:
interpolation to zero at $x=1$.\label{fig:spec.M}}
\figure{Relative contributions of the separate terms in the operator
product expansion
to the spectrum as function of $x$ ($M^2 = 8\mbox{\ GeV}^2$, $m_b =
4.8\mbox{\ GeV}$, $s_0 = 34\mbox{\ GeV}^2$). Solid line:
perturbative
contribution, long dashes: contribution of the quark condensate,
 short
dashes: contribution of the mixed condensate.\label{fig:spec.cont}}
\figure{The spectrum $|V_{ub}|^{-2} d\Gamma/dx$ as function of $x$
 with
$m_b = 4.8\mbox{\ GeV},$ $s_0 = 34\mbox{\ GeV}^2$ and $M^2 =
 8\mbox{\
GeV}^2$. Solid line: the spectrum using (\ref{eq:sr}), long dashes:
 the
same with inclusion of radiative corrections for the free-quark
decay
acc.\ to (\ref{eq:radcorr}).\label{fig:radcorr}}
\figure{$1/\Gamma\: d\Gamma/dx$ as function of $x$ for the free
quark
decay and in different model-calculations. The chosen normalization
emphasizes the differences in shape. ACCM: \cite{accm}, GISW:
\cite{GISW}, RDB: \cite{RDB}, BBD: this paper.\label{fig:models}}

\begin{table}
\caption{Central values for the inclusive semileptonic $B\to
X_u\,e\,\nu$ decay rate from this paper and from different other
calculations.
Free: free quark decay with QCD corrections \cite{accm}, Eq.~(25),
$\alpha_s =0.24,\: m_b=5.0\,\mbox{GeV}$; ACCM: decay including
bound-state corrections \cite{accm}, Eq.~(27), $m_s=0.16 GeV,\;
p_f=0.15 GeV, \alpha_s=0.24$; GISW: \cite{GISW}, Fig.~6;
RDB: \cite{RDB}, Fig.~2, cut-off 1.5~GeV; BBD: this paper. We give
values for the total rate $\Gamma$ in $|V_{ub}|^2 10^{14}\,
\mbox{s}^{-1}$ and for the spectrum integrated over electron
energies
in bins of 0.1~GeV in the rest and in the lab frame. In the lab
frame
the B mesons are assumed to have a momentum of $0.34\,\mbox{GeV}$
and
isotropic angular distribution.}
\begin{tabular}{lllllllll}
Frame & Model &$\Gamma $& 2.1--2.2 & 2.2--2.3 & 2.3--2.4 & 2.4--2.5
 &
2.5--2.6 & 2.6--2.7 \\ \tableline
Rest & BBD  & 0.68 & $0.053$ & $0.050$ & $0.044$ & $0.034$ &
$0.019$ &
$0.002$ \\
  & Free  &0.85 & 0.066 & 0.067 & 0.062 & 0.045 & 0 & 0 \\
  & ACCM  & 0.89 & 0.070 & 0.071 & 0.068 & 0.052 & 0.019 & 0 \\
  & GISW  & -- & 0.033 & 0.030 & 0.023 & 0.013 & 0.003 & 0 \\
  & RDB  & 0.92 & 0.071 & 0.070 & 0.066 & 0.043 & 0.009 & 0 \\
 \tableline
Lab & BBD & 0.68 & $0.052$ & $0.049$ & $0.042$ & $0.032$ & $0.018$ &
$0.008$\\
& Free  & 0.85 & 0.066 & 0.066 & 0.058 & 0.037 & 0.017 & 0.0015 \\
& ACCM   & 0.89 & 0.069 & 0.069 & 0.063 & 0.045 & 0.025  & 0.008 \\
& GISW   &  -- &0.032 & 0.029 & 0.022 & 0.013  & 0.006 & 0.002\\
& RDB  & 0.92 & 0.071 & 0.069 & 0.059 & 0.039 & 0.019 & 0.005

\end{tabular}
\end{table}

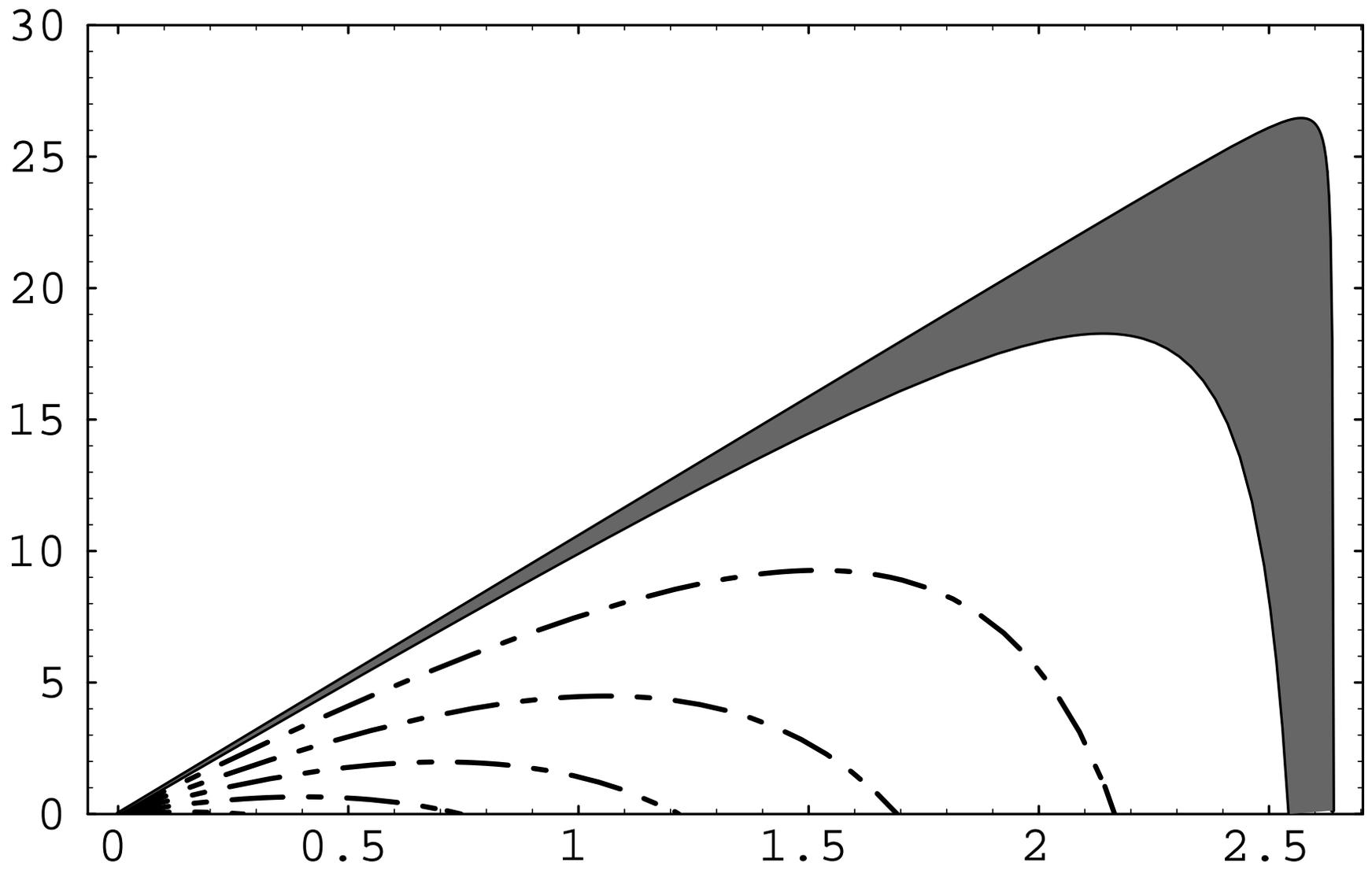
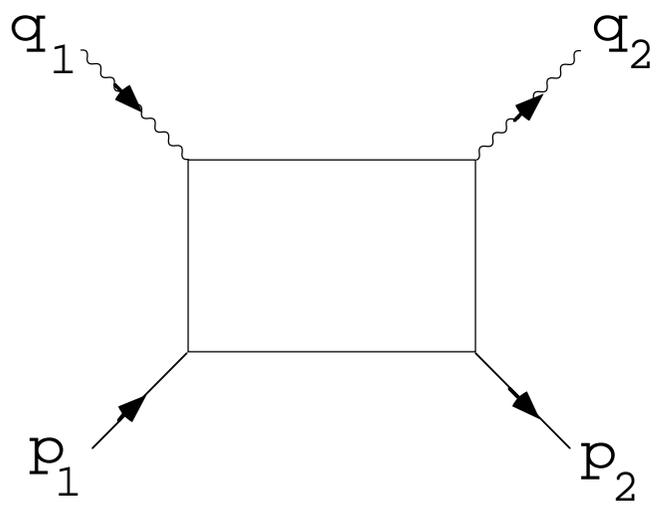
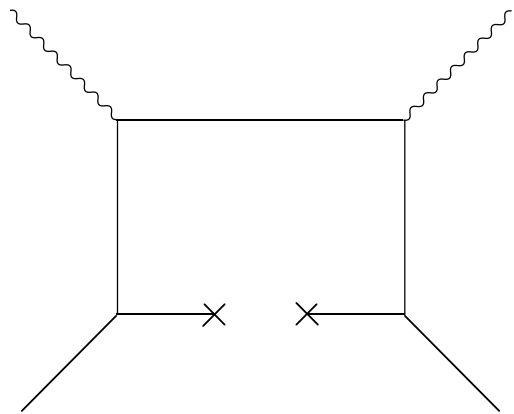
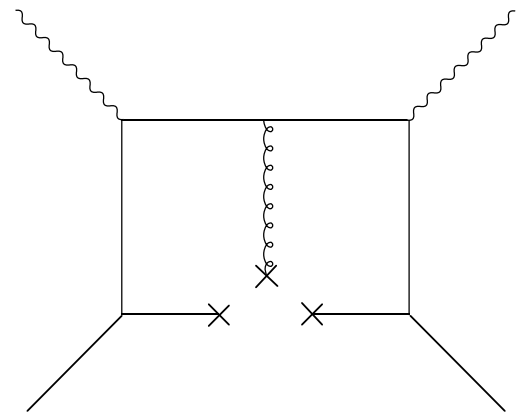
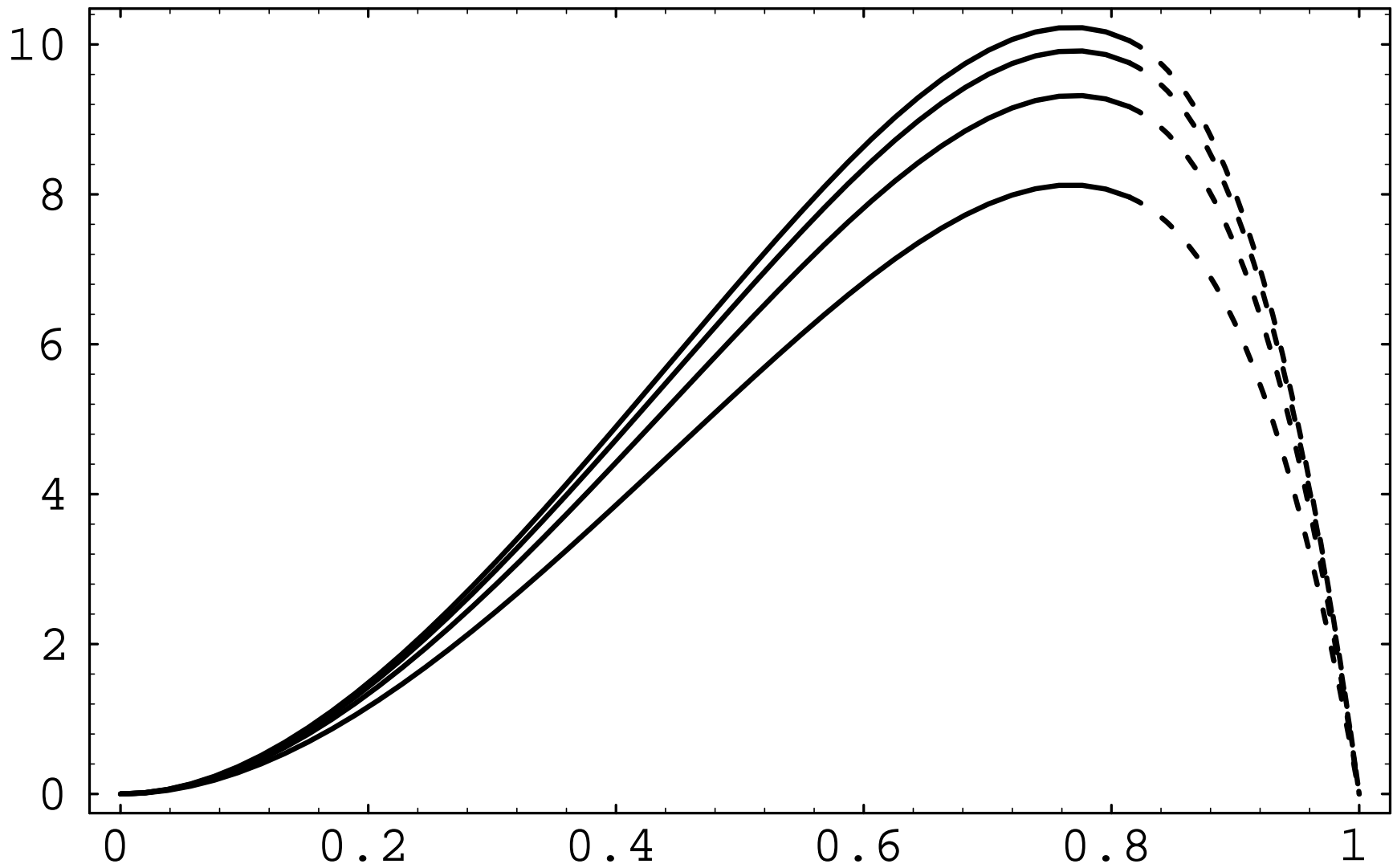
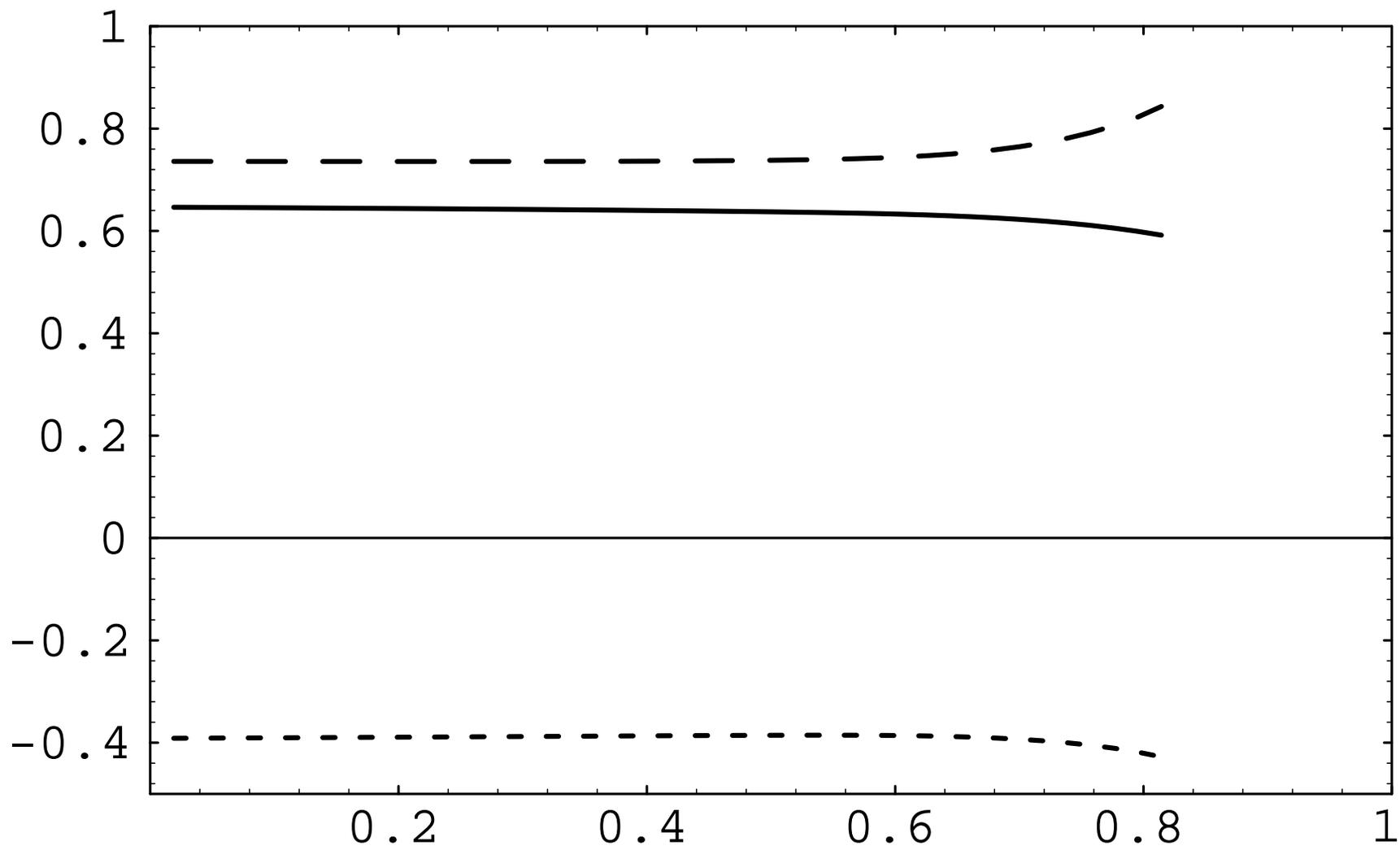
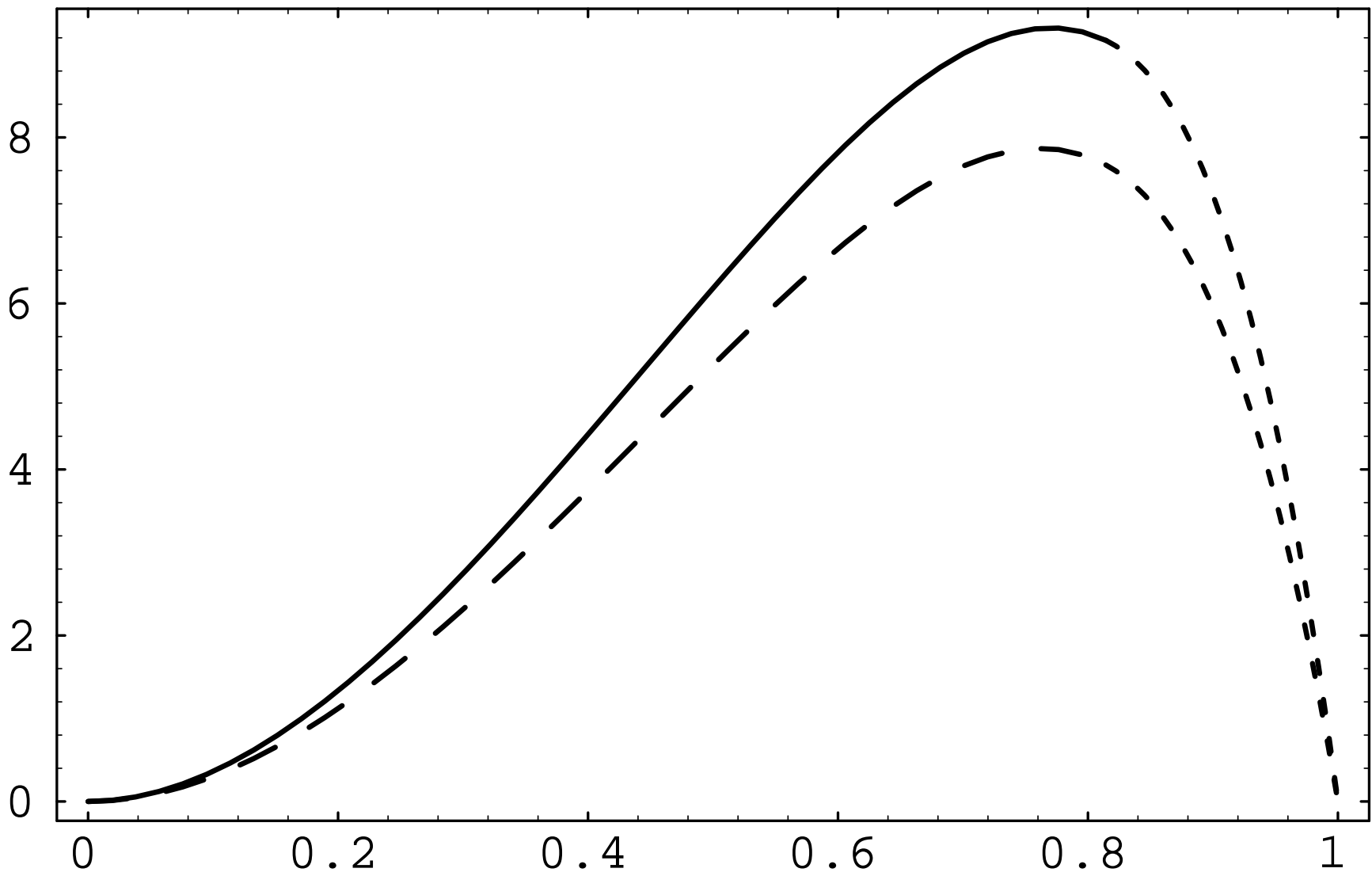
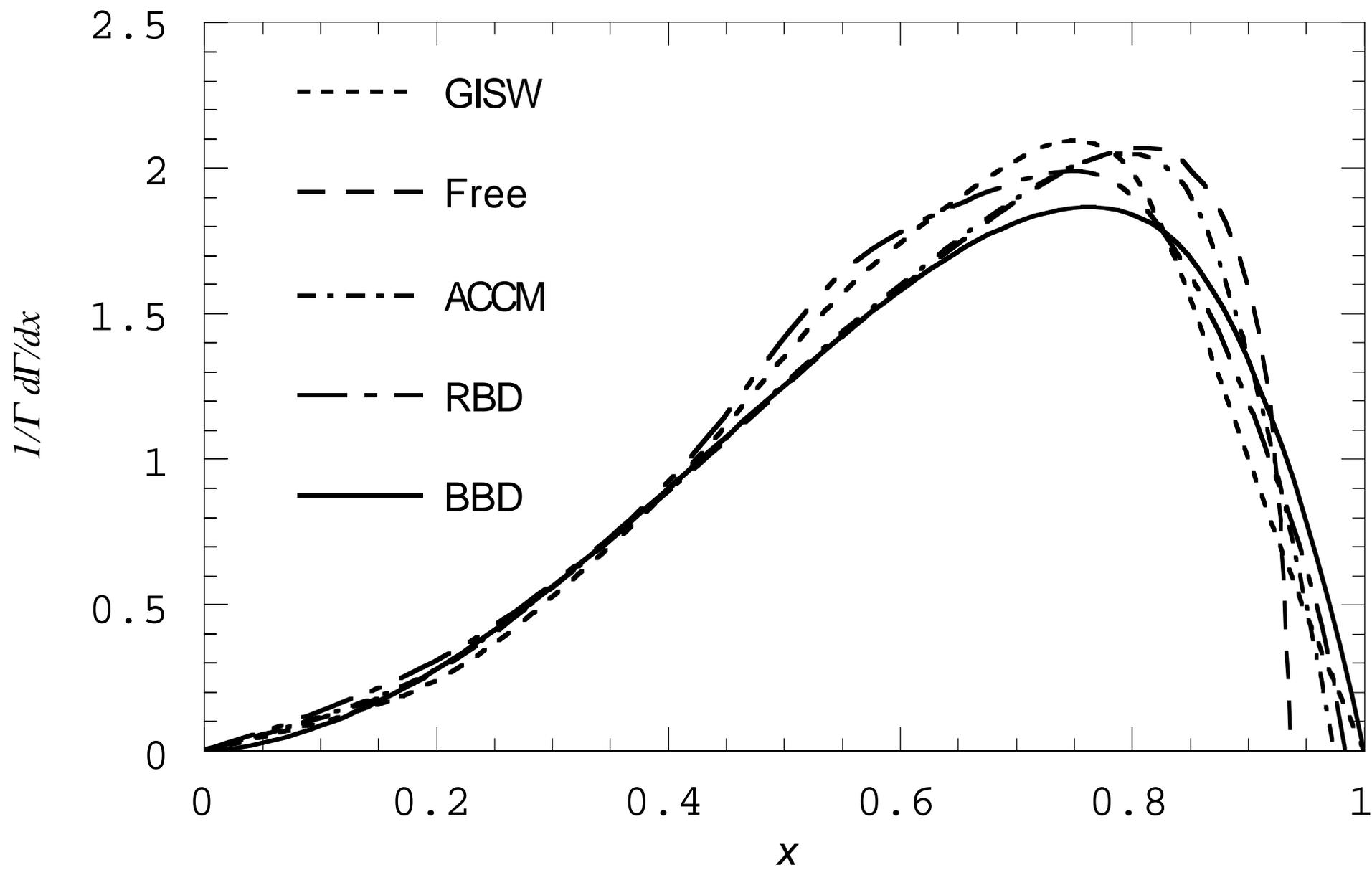
\end{document}